\documentclass{article}

\usepackage{arxiv}

\usepackage[utf8]{inputenc}      
\usepackage[T1]{fontenc}         
\usepackage{mathptmx}            

\usepackage{amssymb, amsmath, amsfonts, bm, mathrsfs} 
\usepackage{nicefrac}            
\usepackage{multirow}            
\usepackage{makecell}            
\usepackage{etoolbox}            

\usepackage{graphicx}            
\usepackage{booktabs, dcolumn}   
\usepackage{microtype}           
\usepackage{xcolor}              

\usepackage[colorlinks=true, linkcolor=blue, citecolor=blue, urlcolor=blue]{hyperref}
\usepackage{url}                 
\usepackage{doi}                 
\usepackage{tabularx} 
\usepackage[labelfont=bf,justification=raggedright,singlelinecheck=false]{caption} 
\usepackage{algorithm}
\usepackage{algpseudocode} 
            
\title{Energy-Conserving Contact Dynamics of Nonspherical Rigid-Body Particles}

\author{
{\hspace{1mm}Haoyuan Shi}\thanks{Corresponding author: haoyuan.shi@pnnl.gov} \\
Physical Sciences Division\\
Pacific Northwest National Laboratory\\
Richland, WA 99354, USA
\And
{\hspace{1mm}Christopher J. Mundy} \\
Physical Sciences Division\\
Pacific Northwest National Laboratory\\
Richland, WA 99354, USA\\
\And
{\hspace{1mm}Gregory K. Schenter} \\
Physical Sciences Division\\
Pacific Northwest National Laboratory\\
Richland, WA 99354, USA
\And
{\hspace{1mm}Jaehun Chun}\thanks{Corresponding author: jaehun.chun@pnnl.gov} \\
Physical Sciences Division\\
Pacific Northwest National Laboratory\\
Richland, WA 99354, USA\\
}

\renewcommand{\shorttitle}{Energy-Conserving Contact Dynamics of Nonspherical Rigid-Body Particles}

\hypersetup{
    pdftitle={Energy-Conserving Contact Dynamics of Nonspherical Rigid-Body Particles},
    pdfsubject={Computational Physics},
    pdfauthor={Haoyuan Shi, Christopher J. Mundy, Gregory K. Schenter, Jaehun Chun},
    pdfkeywords={Rigid-body dynamics, Energy conservation, Particle contact, Anisotropic particles, Molecular simulation}
}

\begin{document}
\maketitle

\begin{abstract}
	Understanding the contact dynamics of nonspherical particles beyond the microscale is crucial for accurately modeling colloidal and granular systems, where shape anisotropy dictates structural organization and transport properties. In this paper, we introduce an energy-conserving contact dynamics framework for arbitrary convex rigid-body particles, integrating vertex–boundary interactions in 2D with vertex–surface and edge–edge detection in 3D. This formulation enables continuous force evaluation and strictly prevents particle overlap while conserving total energy during translational and rotational motion. Simulations of polygonal and polyhedral particles confirm the framework’s stability and demonstrate its capability to capture packing behavior, anisotropic diffusion, and equations of state. The framework establishes a robust and extensible foundation for investigating the nonequilibrium dynamics of complex nonspherical particle systems, with potential applications in colloidal self-assembly, granular flow, and hydrodynamics.
\end{abstract}

\keywords{Rigid-body dynamics, Energy conservation, Particle contact, Anisotropic particles, Molecular simulation}

\section{Introduction}

Particle-based simulations have emerged as a powerful tool for investigating particle-scale behaviors underlying a wide range of phenomena, including granular flow and colloidal self-assembly. Structural and dynamic features inherent from their governing interactions provide fundamental insights into the emergent collective behavior of particulate systems, where the particle shape defines the constituent building blocks that dictate structural organization,\cite{glotzer2007anisotropy,damasceno2012predictive} and the dynamics further determine how such structures evolve and respond under external perturbations.\cite{li2012direction, de2015crystallization} A general approach for modeling shaped particles is through fully atomistic or coarse-grained molecular dynamics (MD) simulations, where complex geometries can be constructed from sub-particles, and the total particle–particle interaction is represented as the sum of interactions between all constituent sub-particles. Furthermore, when only collective properties are of interest, the sub-particles within each particle can be treated as a rigid body, neglecting internal interactions and thereby greatly reducing computational complexity. Nevertheless, key challenges remain. On one hand, fully atomistic models can achieve well-defined interatomic potentials and high accuracy, but computing all relevant interactions for a sufficiently large number of particles at micron-scale lengths is computationally prohibitive. On the other hand, coarse-grained methods, which group atoms from the fully atomistic model to reduce computational cost, cannot accurately capture short-range contact interactions.\cite{noid2023perspective} Specifically, by averaging over atomistic details, coarse-graining modifies the effective surface roughness, thereby altering tangential forces and friction during interparticle contact, which, for example, may dominate momentum transfer and lubrication forces governing the dynamics of nanoparticle assembly.\cite{li2023nanoparticle,heo2024colloidal,krim2024fundamental, shi2025incorporating}

An alternative approach to representing particle shape in simulations is to employ anisotropic interaction potentials with explicit orientational dependence applied to a single particle or point, such as the Gay–Berne potential,\cite{berardi1998gay,berardi2008field}, even though it may not easily generalize to particles with more complex geometries.\cite{nguyen2019aspherical,ramasubramani2020mean} Another method is discrete element models (DEMs), in which particle shape is explicitly defined in the geometry and can be readily implemented within a MD framework for simulations. Interactions between pairs of contact points belonging to two particles are defined and computed to prevent particle overlap. Several open-source packages are available for simulating DEMs, including the Langston model\cite{fraige2008vibration,wang2011particle} implemented in LAMMPS,\cite{thompson2022lammps} as well as the DEM package\cite{spellings2017gpu} and the anisotropic Lennard–Jones (LJ) potential\cite{ramasubramani2020mean} available in HOOMD-Blue,\cite{anderson2020hoomd} each employing distinct schemes for contact-point detection. However, two significant issues remain. First, non-smooth detection of contact points can lead to discontinuities in the computed forces. Second, intersection or penetration may occur because the discrete set of contact points enforces repulsion only locally, leaving distant regions of a particle effectively unconstrained, which can result in unphysical overlaps during large translational or rotational motions. Such issues can significantly impact energy conservation during simulations, which in turn affects the dynamic behavior critical for accurately modeling realistic systems.

In this work, we present a new framework for determining contact in nonspherical rigid-body particles. The approach inherits the linear and frictional contact formulations from the Langston model,\cite{fraige2008vibration,wang2011particle} while introducing a clear and efficient scheme based on vertex–boundary interactions for 2D particles and vertex–surface and edge–edge interactions for 3D particles. Rather than restricting the simulation to a single pair of contact points, our model identifies and includes all possible contact pairs within a specified cutoff distance, including duplicates, thereby maximizing the likelihood of preventing intersection or penetration. Moreover, retaining duplicate contact points is essential for preserving force continuity, as small perturbations can cause slight divergences among duplicates and yield variations in force contributions, leading otherwise to abrupt force discontinuities. The implementation of this framework in LAMMPS\cite{thompson2022lammps} offers scalability, parallel efficiency, and flexibility for incorporating additional non-contact interactions into rigid-body particle simulations. We also conduct extensive tests to assess its energy conservation performance for both 2D and 3D particles and demonstrate its capability to capture a wide range of behaviors in nonspherical particle systems, offering a valuable tool for advancing studies in colloidal and granular dynamics.

This paper is organized as follows. Section~\ref{sec:theory} presents the contact force formulations and the algorithms for detecting contact point pairs in both 2D and 3D. Section~\ref{sec:implementation} describes the implementation of the framework. Section~\ref{sec:2d} examines the energy conservation performance of various 2D particles at different packing fractions and investigates the packing properties at each fraction. Section~\ref{sec:3d} focuses on the energy conservation performance of 3D particles and explores the diffusion behavior, packing properties, and equation of state. Understanding these properties for nonspherical particles is essential for accurately modeling their behavior in diverse applications. Their anisotropic shapes lead to directional diffusion, affecting transport and self-assembly processes.\cite{de2022spiers,li2023nanoparticle} Packing behavior governs structural density, porosity, and mechanical stability, which are critical for material design and granular systems.\cite{van2014understanding, van2015digital, cersonsky2018relevance,tahmasebi2023state} Additionally, the equation of state relates pressure, volume, and temperature, providing key insights into the thermodynamic and collective behavior of dense nonspherical particle assemblies.\cite{glotzer2007anisotropy,agarwal2011mesophase,damasceno2012predictive} Finally, in Section~\ref{sec:con}, we summarize our work, highlight potential applications of the framework, and outline future directions to enhance its versatility and broad applicability.

\section{Theory}
\label{sec:theory}
\subsection{Contact Force Formulation}

For two body particles \((i,j)\) in contact, the contact force on each particle should act at the point on the particle's body where contact occurs. For a 2D body particle, the contact point lies on its boundary, whereas for a 3D body particle, it lies on its surface. The effective separation between two contact points is defined as 
\(\delta_n = d - (R_i + R_j)\), where \(d\) is the distance between the points and \(R_i\) and \(R_j\) denote the skin layer thickness of each particle. For such offset (or filleted) particles, based on and modified from Fraige et al.~\cite{fraige2008vibration} and Wang et al.~\cite{wang2011particle}, the force acting at the contact points between two particles is given by:
\begin{equation}
\label{eq:contact_f}
F_n =
\begin{cases} 
-k_n \delta_n - c_n v_n, & \delta_n \le 0, \\
-k_n \delta_n, & 0< \delta_n \le \frac{k_{na}r_c}{k_n+k_{na}}, \\
k_{na} \delta_n - k_{na} r_c, & \frac{k_{na}r_c}{k_n+k_{na}} < \delta_n \le r_c, \\
0, & \delta_n > r_c,
\end{cases}
\end{equation}
where \(r_c\) is the cutoff distance, \(c_n\) is the damping coefficient associated with the normal relative velocity \(v_n\), and \(k_n\) and \(k_{na}\) are the stiffness coefficients of the repulsive and attractive forces, respectively. The linear contact force provides a computationally efficient approximation of the initial elastic response between particles and can be tuned to match forces from theory, simulations, or experiments, including atomistic potentials or atomic force microscopy measurements.\cite{wang2011particle} In addition to the normal contact force, the tangential force is defined to capture more realistic behaviors, such as in granular systems, where it governs friction, energy dissipation, and the formation of force chains that determine the material’s mechanical response, and is given by:
\begin{equation}
F_t =
\begin{cases}
- c_t v_t, & \delta_n \le 0 \ \text{and}\ |F_t| < \mu |F_n|, \\
- \mu |F_n|, & \delta_n \le 0 \ \text{and}\ |F_t| \ge \mu |F_n|, \\
0, & \delta_n > 0~.
\end{cases}
\end{equation}
where \(c_t\) is the damping coefficient corresponding to the tangential relative velocity \(v_t\), and \(\mu\) denotes the kinetic friction coefficient.

\begin{figure}[htbp]
    \centering
    \includegraphics[width=0.5\linewidth]{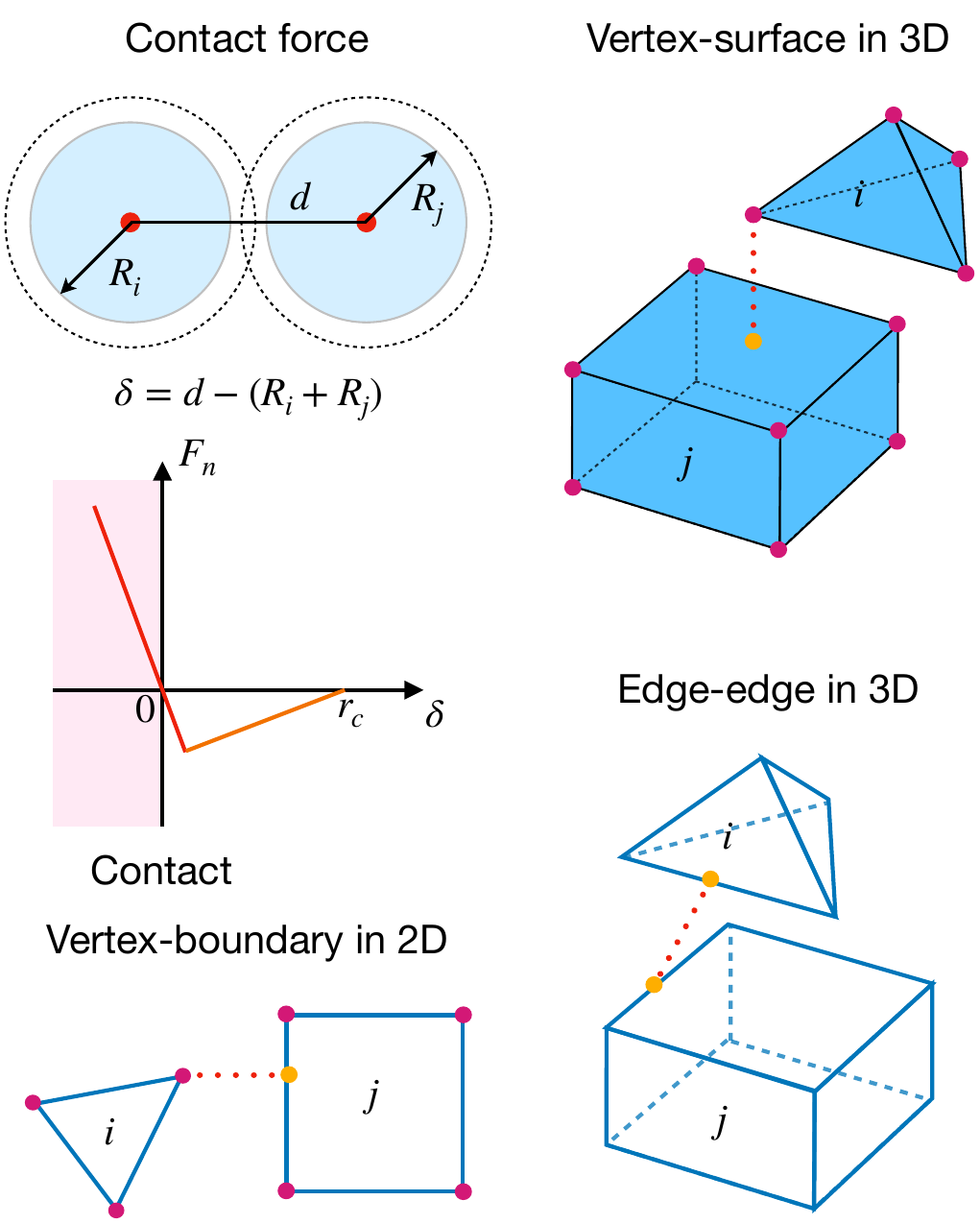}
    \caption{Scheme of contact forces and interactions. The contact force is determined by the interparticle distance: a repulsive force arises when the separation $d$ is smaller than the combined skin layer thickness $(R_i + R_j)$, while the cutoff distance $r_c$ switches on the attractive force for separations beyond $R_i + R_j$. Vertex-boundary interactions in 2D (a triangle and a square) involve 7 ($3+4$) pairs, while in 3D, vertex-surface interactions for a tetrahedron and a cube involve 12 ($4+8$) pairs, and edge-edge interactions involve 72 ($6 \times 12$) pairs.}
    \label{fig:scheme}
\end{figure}

\subsection{2D Body Particle Interaction}
\subsubsection{Vertex–Boundary Interaction}

In two dimensions, the interaction between body particles is governed by the geometry of their boundaries and the positions of their vertices. For a 2D polygon, incorporating a skin layer into the contact force formulation is mathematically equivalent to performing a Minkowski sum of the polygon with a disk of radius equal to the skin thickness, effectively producing an offset geometry. For two interacting polygons \((i,j)\), a contact point pair is defined by a vertex of one particle and the corresponding point on the boundary of the other particle that lies at the shortest distance from this vertex. Note that the polygons must be convex to ensure the nearest point on the boundary is unique. Therefore, if the number of vertices for the two polygons are \(n_i\) and \(n_j\), the algorithm evaluates a total of (\(n_i + n_j\)) vertex–boundary distances. The circular particle can be regarded as a special case represented by a single vertex, where the skin layer thickness corresponds to the radius. In contrast, a 2D rod can be represented as a line segment with two vertices and rounded ends. The pseudocode for detecting vertex–boundary contact points is presented in Algorithm~\ref{alg:VB_code}. Since contact points always occur in pairs and the contact forces act along the same line with equal magnitude and opposite directions, both linear and angular momentum are automatically conserved. With tangential forces, angular momentum is approximately conserved if the contact points are close, i.e., when the skin layer thickness is small relative to the particle size, allowing the forces to be treated as collinear.

\begin{algorithm}[htbp]
\caption{Vertex--Boundary Contact Point Detection}
\label{alg:VB_code}
\begin{algorithmic}[1] 
\State \textbf{Input:} Polygons $P_i, P_j$ with vertices $\mathcal{V}_i, \mathcal{V}_j$; skin thickness $R_i, R_j$; cutoff $r_c$
\State \textbf{Output:} Contact points set $\mathcal{C}_{ij}$
\State Initialize $\mathcal{C}_{ij} \gets \emptyset$
\For{each vertex $\mathbf{v}_k \in \mathcal{V}_i$}
    \State $d_\text{min} \gets \infty$, $\mathbf{p}_\text{min} \gets$ null
    \For{each edge $e_m = (\mathbf{u}_m, \mathbf{u}_{m+1}) \in \partial P_j$}
        \State Project $\mathbf{v}_k$ onto line of $e_m$: $\mathbf{p}_{km}$
        \If{$\mathbf{p}_{km} \notin e_m$}
            \State $\mathbf{p}_{km} \gets$ closest of $\mathbf{u}_m$ or $\mathbf{u}_{m+1}$
        \EndIf
        \State $d_{km} \gets \|\mathbf{v}_k - \mathbf{p}_{km}\|$
        \If{$d_{km} < d_\text{min}$}
            \State $d_\text{min} \gets d_{km}$, $\mathbf{p}_\text{min} \gets \mathbf{p}_{km}$
        \EndIf
    \EndFor
    \If{$d_\text{min} < r_c$}
        \State Add $(\mathbf{v}_k, \mathbf{p}_\text{min})$ to $\mathcal{C}_{ij}$
    \EndIf
\EndFor
\For{each vertex $\mathbf{u}_l \in \mathcal{V}_j$ (roles of $P_i, P_j$ reversed)}
    \State \textit{Repeat above steps}
\EndFor
\State \textbf{Return} $\mathcal{C}_{ij}$
\end{algorithmic}
\end{algorithm}

\subsection{3D Body Particle Interaction}
\subsubsection{Vertex–Surface Interaction}

In three dimensions, vertex–boundary interactions generalize to vertex–surface interactions, where a contact point consists of a vertex of one polyhedron and its nearest point on the surface of another. Still, the polyhedron must be convex to ensure a unique shortest distance. Spheres and 3D rods represent special cases with one and two vertices, respectively, extended by a skin layer. The pseudocode for detecting vertex–surface contact points is presented in Algorithm~\ref{alg:VS_code}. 

\begin{algorithm}[htbp]
\caption{Vertex--Surface Contact Point Detection in 3D}
\label{alg:VS_code}
\begin{algorithmic}[1]
\State \textbf{Input:} Polyhedra $P_i, P_j$ with vertices $\mathcal{V}_i, \mathcal{V}_j$ and faces $\mathcal{F}_i, \mathcal{F}_j$; skin thickness $R_i, R_j$; cutoff $r_c$
\State \textbf{Output:} Contact points set $\mathcal{C}_{ij}$
\State Initialize $\mathcal{C}_{ij} \gets \emptyset$
\For{each vertex $\mathbf{v}_k \in \mathcal{V}_i$}
    \State $d_\text{min} \gets \infty$, $\mathbf{p}_\text{min} \gets$ null
    \For{each face $f_m \in \mathcal{F}_j$}
        \State Project $\mathbf{v}_k$ onto the plane of $f_m$: $\mathbf{p}_{km}$
        \If{$\mathbf{p}_{km} \notin f_m$}
            \State $\mathbf{p}_{km} \gets$ closest point on the boundary of $f_m$ (vertex--boundary)
        \EndIf
        \State $d_{km} \gets \|\mathbf{v}_k - \mathbf{p}_{km}\|$
        \If{$d_{km} < d_\text{min}$}
            \State $d_\text{min} \gets d_{km}$, $\mathbf{p}_\text{min} \gets \mathbf{p}_{km}$
        \EndIf
    \EndFor
    \If{$d_\text{min} < r_c$}
        \State Add $(\mathbf{v}_k, \mathbf{p}_\text{min})$ to $\mathcal{C}_{ij}$
    \EndIf
\EndFor
\For{each vertex $\mathbf{u}_l \in \mathcal{V}_j$ (roles of $P_i, P_j$ reversed)}
    \State \textit{Repeat above steps}
\EndFor
\State \textbf{Return} $\mathcal{C}_{ij}$
\end{algorithmic}
\end{algorithm}

\subsubsection{Edge–Edge Interaction}

Vertex–surface interactions in 3D are not sufficient to fully describe contacts between two polyhedra, as they can fail to detect contacts where two edges intersect or penetrate each other. Therefore, an edge–edge interaction algorithm is employed, considering all \(m_i \times m_j\) edge pairs of the respective polyhedra, with contact points defined as the closest points on each edge. The pseudocode is presented in Algorithm~\ref{alg:EE_code}. It is worth noting that different edges can share a common vertex, so edge–edge interactions may yield identical contact points when one or both points coincide with a vertex. These contacts may also overlap with those detected in vertex–surface interactions. Retaining such duplicate contact pairs is not only allowed but essential, since as the polyhedra evolve, these initially coincident pairs can diverge into distinct contact points. Ignoring them would lead to discontinuities in the computed forces.

\begin{algorithm}[htbp]
\caption{Edge--Edge Contact Point Detection in 3D}
\label{alg:EE_code}
\begin{algorithmic}[1]
\State \textbf{Input:} Polyhedra $P_i, P_j$ with vertices $\mathcal{V}_i, \mathcal{V}_j$ and edges $\mathcal{E}_i, \mathcal{E}_j$; skin thickness $R_i, R_j$; cutoff $r_c$
\State \textbf{Output:} Contact points set $\mathcal{C}_{ij}$
\State Initialize $\mathcal{C}_{ij} \gets \emptyset$
\For{each edge $e_a = (\mathbf{u}_1, \mathbf{u}_2) \in \mathcal{E}_i$}
    \For{each edge $e_b = (\mathbf{v}_1, \mathbf{v}_2) \in \mathcal{E}_j$}
        \If{$\mathbf{e}_a \nparallel \mathbf{e}_b$}
            \State Find $\mathbf{p}_a = \mathbf{u}_1 + t (\mathbf{u}_2-\mathbf{u}_1), \;
                   \mathbf{p}_b = \mathbf{v}_1 + s (\mathbf{v}_2-\mathbf{v}_1)$
            \State $t,s \in [0,1]$ such that $\|\mathbf{p}_a-\mathbf{p}_b\|$ is minimized
        \Else
            \State $\mathbf{p}_a, \mathbf{p}_b \gets$ midpoint of overlap or nearest endpoints
        \EndIf
        \State $d_{ab} \gets \|\mathbf{p}_a - \mathbf{p}_b\|$
        \If{$d_{ab} < r_c$}
            \State Add $(\mathbf{p}_a, \mathbf{p}_b)$ to $\mathcal{C}_{ij}$
        \EndIf
    \EndFor
\EndFor
\State \textbf{Return} $\mathcal{C}_{ij}$
\end{algorithmic}
\end{algorithm}

\section{Implementation}
\label{sec:implementation}
The proposed contact detection algorithms are implemented in the Large-scale Atomic/Molecular Massively Parallel Simulator (LAMMPS),\cite{thompson2022lammps} by extending the body-particle package~\cite{fraige2008vibration, wang2011particle} (\url{https://docs.lammps.org/Howto_body.html}), leveraging LAMMPS’s scalability, parallel efficiency, and ability to incorporate additional non-contact interactions into body-particle simulations. The input data file containing the shape information and the LAMMPS script remain the same as in the original version. The \textit{coxeter} Python package~\cite{ramasubramani2021coxeter} provides tools to easily generate shape information that can be directly used in the input data file. The LAMMPS dump file can contain coordinates and quaternions for each particle, which can then be converted into a GSD file~\cite{anderson2020hoomd} and visualized using OVITO.\cite{stukowski2009visualization} In this work, all simulations employ the Lennard–Jones (LJ) reduced unit system, with variables denoted by a superscript $^*$ as reduced quantities, resulting in a fully nondimensional representation.

\section{2D Body Particle Simulation}
\label{sec:2d}
\subsection{Energy Conservation}

To assess the physical validity of the 2D body-particle simulations, we examine energy conservation for systems composed of polygons, each with a uniform effective area \(s_{o}^*\), distributed within a simulation domain of area \(S^*\). The packing fraction is defined as \(\eta = s_{o}^* N / S^*\), where \(N\) is the total number of particles. Here, only the conservative repulsive force is considered (\(F = -k_n^* \delta_n^*\)), with a stiffness of \(k_n^* = 300\), and a skin layer thickness of \(R^*=0.15\). The effective area \(s_{o}^*\) is then taken as the Minkowski sum of the unit-area polygon and a disk of radius \(R^*\). Each simulation consists of a sequence of NPT ensembles, maintained with a Nose–Hoover thermostat and barostat~\cite{martyna1994constant} to control the system at different pressures, each run for 200,000 steps (\(20\tau\) with a timestep \(\Delta t=0.0001\tau\)), followed by an 200,000 steps NVE ensemble to monitor energy conservation. The system temperature is set to \(T^* = 1\) with a thermostat damping time of 100 timesteps, and the target pressures are \(P^* = 1, 3, 6, 7, 8, 10\) with a barostat damping time of 1000 timesteps. Each simulation box contains 196 particles with initially random positions and orientations.

\begin{figure}[htbp]
    \centering
    \includegraphics[width=\linewidth]{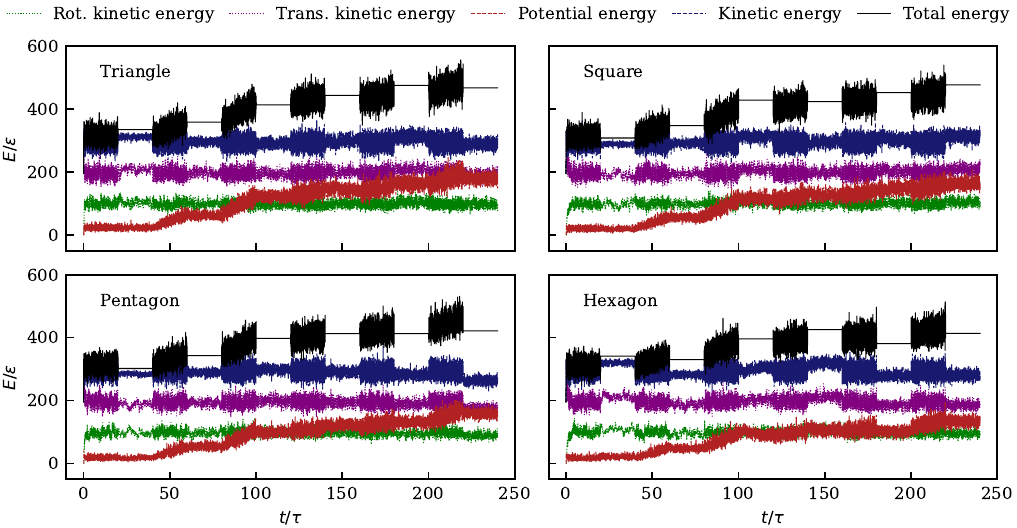}
    \caption{Energy evolution for various 2D shapes, including triangles, squares, pentagons, and hexagons. The simulation cycles through NPT for $20\tau$ and NVE for $20\tau$ at $T^* = 1$ and target pressures $P^* = 1, 3, 6, 7, 8,$ and $10$.}
    \label{fig:2d_eng}
\end{figure}

Figure~\ref{fig:2d_eng} shows the evolution of the energy $E$ over time $t$, demonstrating perfect energy conservation in each NVE ensemble, with no observable drift in either kinetic or potential energy. Furthermore, Table~\ref{tab:shape_results} presents the packing fraction, relative energy differences, average temperature, and average pressure for the various particle shapes across different NVE ensembles, with the energy differences on the order of $10^{-6}$.

\begin{table}[htbp]
    \centering
    \renewcommand{\arraystretch}{1.2} 
    \setlength{\tabcolsep}{7pt} 
    \caption{Energy difference $\Delta E^*/E^*$ for triangular, square, pentagonal, and hexagonal particles at various packing fractions $\eta$ in NVE ensembles following NPT ensembles at different target pressures (Fig.~\ref{fig:2d_eng}), together with the average temperature $T^*$ and average pressure $P^*$.}
    \vspace{0.3cm}
    \begin{tabular}{lcccc}
        \toprule
        \textbf{Shape} & \(\eta\) & \(\Delta E^*/E^*\) & \(T^*\) & \(P^*\) \\
        \midrule
        \multirow{6}{*}{Triangle}
         & 0.43 & \(-2.40\times 10^{-7}\)  & \(1.06 \pm 0.02\) & \(0.98 \pm 0.10\) \\
         & 0.65 & \(-3.31\times 10^{-6}\)  & \(1.01 \pm 0.03\) & \(2.86 \pm 0.20\) \\
         & 0.80 & \(1.62\times 10^{-6}\) & \(0.99 \pm 0.03\) & \(6.05 \pm 0.29\) \\
         & 0.85 & \(-7.03\times 10^{-7}\)  & \(1.02 \pm 0.04\) & \(7.21 \pm 0.34\) \\
         & 0.87 & \(-6.53\times 10^{-7}\)  & \(1.07 \pm 0.04\) & \(8.20 \pm 0.35\) \\
         & 0.91 & \(5.83\times 10^{-7}\) & \(0.98 \pm 0.04\) & \(9.51 \pm 0.36\) \\
        \midrule
        \multirow{6}{*}{Square}
         & 0.45 & \(1.16\times 10^{-7}\) & \(0.99 \pm 0.02\) & \(1.00 \pm 0.11\) \\
         & 0.65 & \(1.10\times 10^{-7}\) & \(1.00 \pm 0.03\) & \(2.90 \pm 0.22\) \\
         & 0.77 & \(9.44\times 10^{-7}\) & \(1.07 \pm 0.03\) & \(5.98 \pm 0.30\) \\
         & 0.81 & \(1.69\times 10^{-6}\) & \(1.01 \pm 0.04\) & \(6.93 \pm 0.36\) \\
         & 0.86 & \(1.13\times 10^{-6}\) & \(1.04 \pm 0.04\) & \(8.44 \pm 0.40\) \\
         & 0.88 & \(1.72\times 10^{-6}\) & \(1.07 \pm 0.04\) & \(9.68 \pm 0.39\) \\
        \midrule
        \multirow{6}{*}{Pentagon}
         & 0.44 & \(9.17\times 10^{-7}\) & \(0.97 \pm 0.02\) & \(0.92 \pm 0.10\) \\
         & 0.66 & \(-2.89\times 10^{-7}\)  & \(0.99 \pm 0.03\) & \(3.04 \pm 0.22\) \\
         & 0.76 & \(-1.11\times 10^{-6}\)  & \(1.02 \pm 0.03\) & \(5.61 \pm 0.30\) \\
         & 0.82 & \(1.25\times 10^{-6}\) & \(1.00 \pm 0.03\) & \(7.33 \pm 0.35\) \\
         & 0.84 & \(-7.40\times 10^{-7}\)  & \(0.96 \pm 0.03\) & \(7.97 \pm 0.36\) \\
         & 0.87 & \(-1.74\times 10^{-6}\)  & \(0.90 \pm 0.03\) & \(9.77 \pm 0.35\) \\
        \midrule
        \multirow{6}{*}{Hexagon}
         & 0.46 & \(-6.00\times 10^{-7}\)  & \(1.09 \pm 0.02\) & \(1.17 \pm 0.12\) \\
         & 0.66 & \(-6.28\times 10^{-7}\)  & \(0.96 \pm 0.02\) & \(2.95 \pm 0.22\) \\
         & 0.79 & \(-2.97\times 10^{-7}\)  & \(1.03 \pm 0.04\) & \(6.06 \pm 0.42\) \\
         & 0.83 & \(-8.27\times 10^{-7}\)  & \(1.09 \pm 0.04\) & \(7.07 \pm 0.40\) \\
         & 0.85 & \(-7.37\times 10^{-7}\)  & \(0.95 \pm 0.03\) & \(7.32 \pm 0.37\) \\
         & 0.89 & \(-1.97\times 10^{-6}\)  & \(0.96 \pm 0.03\) & \(9.79 \pm 0.42\) 
         \\
         \bottomrule
    \end{tabular}
    \label{tab:shape_results}
\end{table}

\subsection{Packing Properties}

\begin{figure}[htbp]
    \centering
    \includegraphics[width=\linewidth]{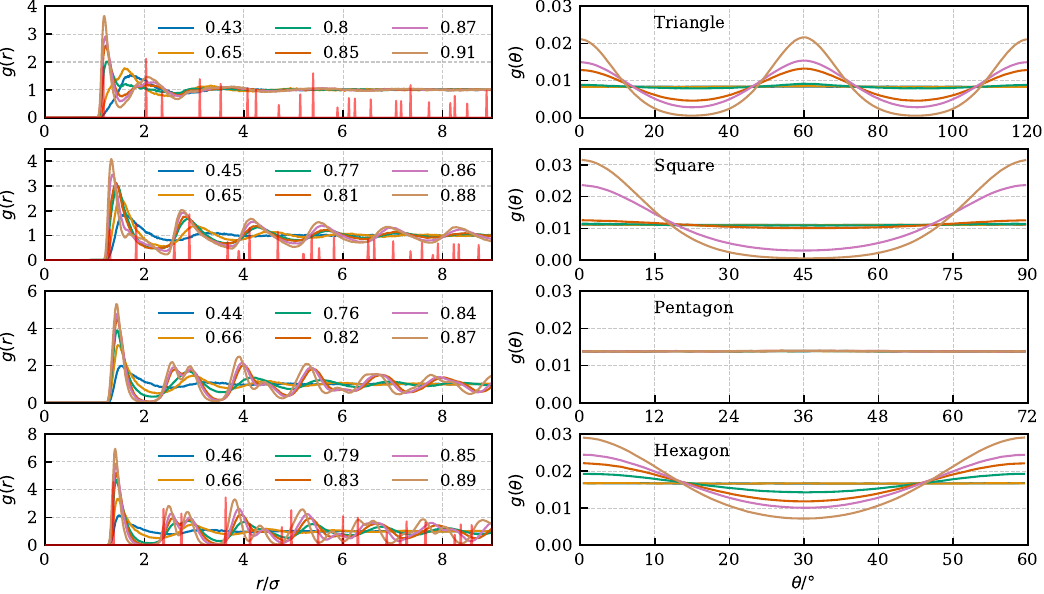}
    \caption{Averaged radial distribution functions $g(r)$ and angular distribution functions $g(\theta)$ for 2D particles with triangular, square, pentagonal, and hexagonal shapes at various packing fractions \(\eta\). Scaled perfect tessellation $g(r)$ curves are shown in red for comparison.}
    \label{fig:2d_pack}
\end{figure}

Figure~\ref{fig:2d_pack} presents the radial distribution functions, $g(r)$, and angular distribution functions, $g(\theta)$, for the 2D particles, averaged over 2,000 frames from each NVE ensemble shown in Fig.~\ref{fig:2d_eng} at different packing densities. As expected, increasing the packing fraction results in more well-ordered configurations, approaching the perfect tessellation for triangles, squares, and hexagons. In the case of pentagons, however, geometric frustration prevents $g(\theta)$ from developing, even though $g(r)$ exhibits a pattern similar to that of hexagons. Note that, although packing configurations can also be obtained using hard particle Monte Carlo (HPMC) to prevent overlaps, our method, which exhibits exact energy conservation in the NVE ensemble, enables the investigation of non-equilibrium phenomena, kinetics, and transport properties that require fully resolved particle dynamics and are inaccessible to standard MC techniques. One example is the precise characterization of local packing defects and their temporal evolution, capturing the dynamics of particle collisions and rearrangements.\cite{torquato2010jammed}

\section{3D Body Particle Simulation}
\label{sec:3d}
\subsection{Energy Conservation}

To assess the energy conservation in the 3D body-particle simulation, we consider a system composed of rods and a mixture of polyhedra, including cubes, large cubes, tetrahedra, and hexagonal prisms, with \(N=200\) of each shape and unit mass. The volumes of the cube, tetrahedron, and hexagonal prism are 1, while the volume of the large cube is 8. The effective volume \(v_{o}^*\) is then defined as the Minkowski sum of the polyhedron and a sphere of radius \(R^* = 0.15\). Each rod consists of a line segment of length 4 with a skin layer thickness of 0.5. The packing fraction is defined as \(\eta = (\sum v_{o}^*) N/V^*\), where \(\sum v_{o}^*\) denotes the total effective volume of the five particle types, and \(V^*\) is the system volume. The stiffness of the conservative repulsive force is \(k_n^* = 500\). Each simulation consists of a sequence of NPT ensembles, as in the 2D particle case, controlling the system at different pressures. Each NPT run is performed for 200,000 steps (\(20\tau\) with a timestep \(\Delta t=0.0001\tau\)), followed by 200,000 steps (\(20\tau\)) of NVT ensembles, and finally 400,000 steps (\(40\tau\)) of NVE ensembles to monitor energy conservation. The system temperature is set to \(T^* = 1.5\) with a thermostat damping time of 100 timesteps, and the target pressures are \(P^* = 0.2, 1, 2, 4, 10\) with a barostat damping time of 1000 timesteps.

\begin{figure}[htbp]
    \centering
    \includegraphics{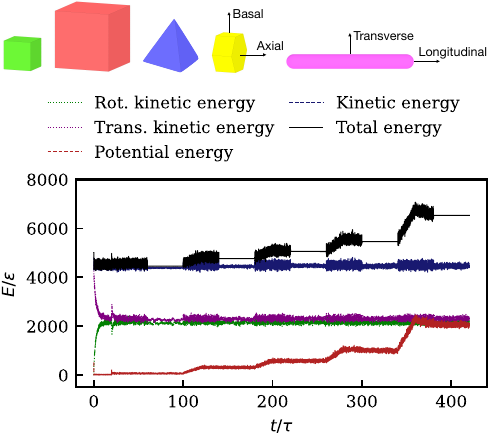}
    \caption{Energy evolution for a mixed system of 3D particles, including cubes, large cubes, tetrahedra, hexagonal prisms, and rods, with 200 of each shape. The simulation cycles through NPT for $20\tau$, NVT for $20\tau$, and NVE for $40\tau$ at target pressures $P^* = 0.2, 1, 2, 4,$ and $10$, and temperature $T^* = 1.5$.}
    \label{fig:diff_eng}
\end{figure}

Figure~\ref{fig:diff_eng} shows the energy evolution over time, and Table~\ref{tab:energy_results} summarizes the packing fraction, relative energy differences, average temperature, and average pressure for the different NVE ensembles. The results indicate no observable drift in either kinetic or potential energy, with energy differences on the order of \(10^{-5}\) even at the largest packing fraction.

\begin{table}[htbp]
    \centering
    \renewcommand{\arraystretch}{1.2} 
    \setlength{\tabcolsep}{7pt} 
    \caption{Energy differences $\Delta E^*/E^*$ at various packing fractions $\eta$ in NVE ensembles following NPT and NVT ensembles, along with the corresponding average temperature $T^*$ and pressure $P^*$ (see Fig.~\ref{fig:diff_eng}).}
    \vspace{0.3cm}
    \begin{tabular}{cccc}
        \toprule
        \(\eta\) & \(\Delta E^*/E^*\) & \(T^*\) & \(P^*\) \\
        \midrule
        0.16 & \(1.9\times10^{-6}\) & \(1.46 \pm 0.004\) & \(0.19 \pm 0.01\) \\
        0.36 & \(-1.2\times10^{-6}\)  & \(1.48 \pm 0.008\) & \(0.98 \pm 0.04\) \\
        0.46 & \(-9.6\times10^{-6}\)  & \(1.49 \pm 0.011\) & \(1.98 \pm 0.07\) \\
        0.57 & \(-9.6\times10^{-6}\)  & \(1.49 \pm 0.013\) & \(3.84 \pm 0.11\) \\
        0.72 & \(1.7\times10^{-5}\) & \(1.49 \pm 0.017\) & \(9.12 \pm 0.19\) \\
		\bottomrule
    \end{tabular}
    \label{tab:energy_results}
\end{table}

\subsection{Particle Diffusion}

To examine the influence of particle geometry on microscopic transport, we evaluate the particle diffusion at various packing fractions under NVE conditions, as summarized in Fig.~\ref{fig:diff_eng} and Table~\ref{tab:energy_results}. Diffusion serves as a fundamental dynamical observable that connects molecular-scale motion to macroscopic transport behavior. In systems composed of shape-anisotropic particles, variations in surface topology and hydrodynamic motion give rise to distinct translational and rotational mobilities, offering valuable insight into how particle geometry governs momentum transfer at the nanoscale.\cite{li2023nanoparticle} In addition to the laboratory-frame translational diffusion coefficient $D_{\mathrm{T}}^{\mathrm{Lab}}$, the body-frame translational $D_{\mathrm{T}}^{\mathrm{Body}}$ and rotational diffusion coefficients $D_{\mathrm{R}}^{\mathrm{Body}}$ are obtained by projecting the linear and angular velocity vectors onto the principal axes of each particle. The body-frame diffusivity specifically captures the translational and rotational motion arising from anisotropic dynamics in particles with axial symmetry, such as rods and hexagonal prisms.

All diffusion coefficients are calculated from the Einstein relation over the last $20\tau$ of each NVE ensemble, with the linear fit applied for $t > 2\tau$ to exclude the ballistic regime. In discrete form, the general expression is
\begin{equation}
    D_{\alpha} = \frac{1}{2 d_{\alpha}} \frac{d}{dt}
     \left\langle \left| \Delta \boldsymbol{\xi}_{\alpha,j}(t) \right|^2 \right\rangle,
    \label{eq:einstein_general}
\end{equation}
where $\alpha$ denotes the type of motion, $d_{\alpha}$ is the number of degrees of freedom, $\Delta \boldsymbol{\xi}_{\alpha,j}(t)$ is the displacement of particle $j$ associated with motion $\alpha$, and the average $\langle \cdot \rangle$ is taken over all $N$ particles. Specifically, the displacements are computed as
\begin{align}
    \Delta \boldsymbol{\xi}^{\mathrm{Lab}}_{\mathrm{T},j}(t) &= \sum_{i=0}^{t/\Delta t-1} \mathbf{v}_j(t_i) \, \Delta t, \\
    \Delta \boldsymbol{\xi}^{\mathrm{Body}}_{\mathrm{T},j}(t) &= 
        \sum_{i=0}^{t/\Delta t-1} \mathbf{R}_j^\mathrm{T}(t_i) \, \mathbf{v}_j(t_i) \, \Delta t, \\
    \Delta \boldsymbol{\xi}^{\mathrm{Body}}_{\mathrm{R},j}(t) &= 
        \sum_{i=0}^{t/\Delta t-1} \mathbf{R}_j^\mathrm{T}(t_i) \, \boldsymbol{\omega}_j(t_i) \, \Delta t,
\end{align}
where $\mathbf{v}_j(t_i)$ and $\boldsymbol{\omega}_j(t_i)$ are the linear and angular velocities of particle $j$ in the laboratory frame at time step $i$, and $\mathbf{R}_j(t_i)$ is the rotation matrix that transforms vectors from the laboratory frame to the particle's body frame. This matrix can be computed from the particle's quaternion $\mathbf{q} = (q_0, q_1, q_2, q_3)$ as
\begin{equation}
\mathbf{R} =
\begin{bmatrix}
1 - 2(q_2^2 + q_3^2) & 2(q_1 q_2 - q_3 q_0) & 2(q_1 q_3 + q_2 q_0) \\
2(q_1 q_2 + q_3 q_0) & 1 - 2(q_1^2 + q_3^2) & 2(q_2 q_3 - q_1 q_0) \\
2(q_1 q_3 - q_2 q_0) & 2(q_2 q_3 + q_1 q_0) & 1 - 2(q_1^2 + q_2^2)
\end{bmatrix}.
\end{equation}
Due to the anisotropic shapes, the body-frame diffusion is evaluated separately along the basal and axial axes for the hexagonal prisms, and along the transverse and longitudinal axes for the rods. Since the rod is essentially a line segment with a skin layer, it cannot undergo rotation about its longitudinal axis.

\begin{figure}[htbp]
    \centering
    \includegraphics{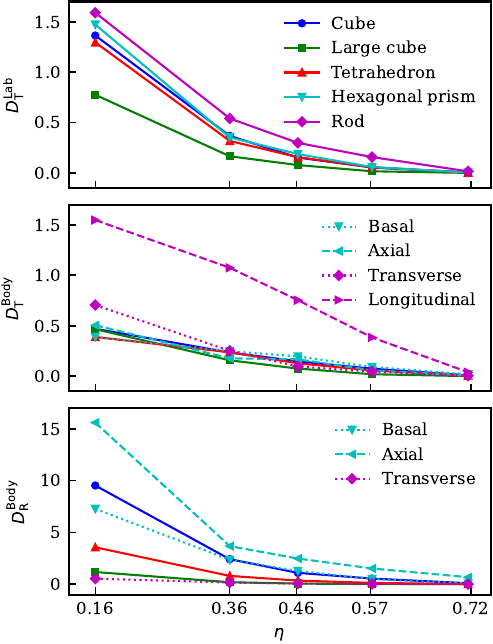}
    \caption{Laboratory-frame translational diffusion, $D_{\mathrm{T}}^{\mathrm{Lab}}$, and body-frame translational, $D_{\mathrm{T}}^{\mathrm{Body}}$, and rotational, $D_{\mathrm{R}}^{\mathrm{Body}}$, diffusion as a function of the total packing fraction $\eta$ for cubes, large cubes, tetrahedra, hexagonal prisms, and rods (see Fig.~\ref{fig:diff_eng}). Body-frame diffusion of anisotropic particles is evaluated along the basal and axial axes for hexagonal prisms and along the transverse and longitudinal axes for rods.}
    \label{fig:3d_diff}
\end{figure}

Figure~\ref{fig:3d_diff} shows the diffusion coefficients for five particle types at five different packing fractions. All diffusion coefficients decrease with increasing packing fraction, with the smallest $D_{\mathrm{T}}^{\mathrm{Lab}}$ observed for the large cubes due to their larger volume, which restricts their mobility. Furthermore, rods exhibit a larger $D_{\mathrm{T}}^{\mathrm{Body}}$ along their longitudinal direction compared to other particle shapes, whereas hexagonal prisms display an enhanced $D_{\mathrm{R}}^{\mathrm{Body}}$ about their axial axis due to their nearly circular basal geometry. These results demonstrate that particle shape and symmetry critically influence transport properties, offering insight into how anisotropic geometries affect diffusion, crowding, and collective dynamics in dense particulate systems.

\subsection{Packing Properties}

As discussed in the 2D packing properties above, the particle dynamics simulations naturally capture the system’s temporal evolution, enabling continuous tracking of particle trajectories during the process. As the packing fraction increases, spatial confinement promotes partial local ordering that resembles Voronoi tessellations,\cite{damasceno2012predictive} giving rise to crystalline assemblies of cubes in simple cubic (SC), rhombic dodecahedra in FCC, and truncated octahedra in BCC. However, in dynamic simulations, system parameters significantly affect the assembly process and the resulting configurations, with the system potentially becoming trapped in local minima when energy barriers are high. Here, we perform simulations with 1000 particles for cubes, rhombic dodecahedra, and truncated octahedra. Each system undergoes an NPT simulation with $P^*$ increased from 1 to 20 over 5,000,000 steps ($500\tau$) at $T^* = 1.5$, followed by 200,000 steps ($20\tau$) of NPT at $P^* = 20$, and concludes with 200,000 steps ($20\tau$) in the NVE ensemble for production. All particles have unit volume and mass, a skin layer thickness of $R^* = 0.15$, and an effective volume defined by the Minkowski sum.

\begin{figure}[htbp]
    \centering
    \includegraphics[width=\linewidth]{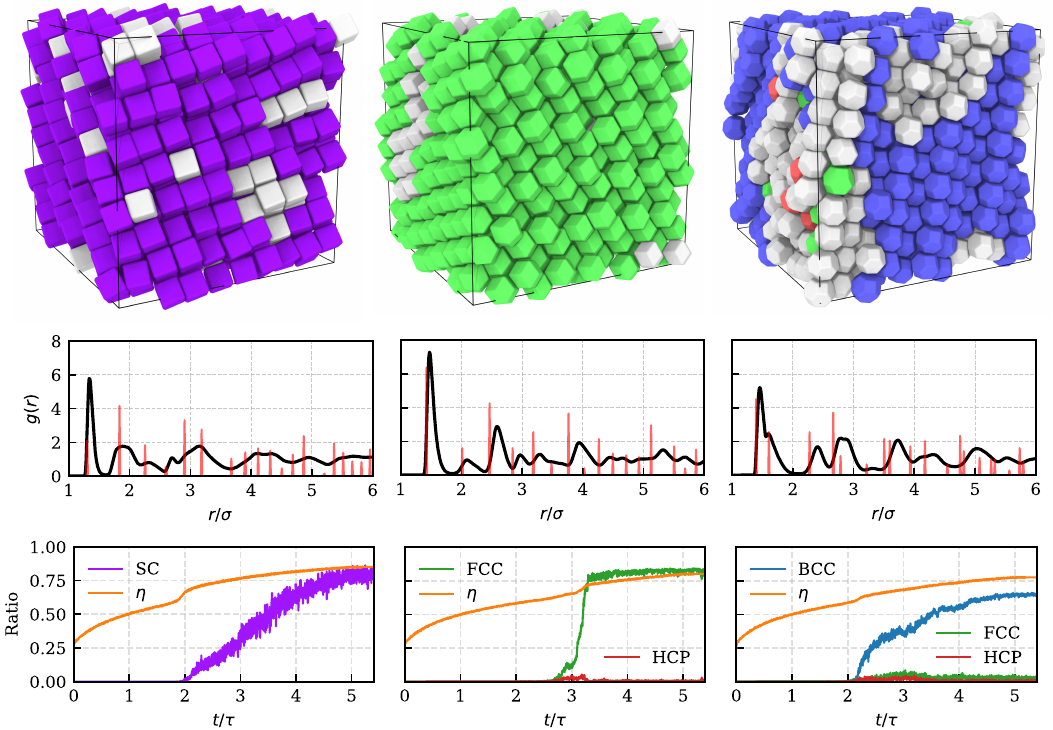}
    \caption{Packed crystal structures of cubes in a simple cubic (SC) lattice (left), rhombic dodecahedra in a FCC lattice with few HCP particles (middle), and truncated octahedra in a BCC lattice with few FCC and HCP particles (right). Final configurations are visualized using Ovito~\cite{stukowski2009visualization} and colored via the built-in polyhedral template matching (PTM) method.\cite{larsen2016robust} The average $g(r)$ in the NVE ensemble and the scaled perfect tessellation $g(r)$ are shown, with the latter plotted in red for comparison. Packing fraction $\eta$ and crystallization ratio are included to illustrate the evolution of dynamics during the simulations.}
    \label{fig:3d_pack}
\end{figure}

Figure~\ref{fig:3d_pack} illustrates the final configurations of the three particle types, the averaged $g(r)$ compared with the ideal perfect tessellation, and the time evolution of the packing fraction $\eta$ and crystallization ratio. The well-ordered configurations closely approach the perfect tessellation. Using the polyhedral template matching (PTM) method~\cite{larsen2016robust} with an RMSD cutoff of 0.1 for crystal structure identification, cubes and rhombic dodecahedra form SC and FCC structures with final fractions exceeding 0.75 at a packing fraction of about 0.8, and the rhombic dodecahedra show a sharp increase in FCC crystallinity around $3\tau$. Moreover, a small fraction of HCP structures forms during the FCC crystallization of rhombic dodecahedra. On the other hand, truncated octahedra attain a slightly reduced BCC crystallinity of approximately 0.65, with a small fraction of FCC and HCP structures. This may be attributed to changes in the equilateral properties of truncated octahedra after adding the skin layer, caused by the heterogeneous face shapes and different dihedral angles.

\subsection{Equation of State}

We investigate the pressure as a function of packing fraction for spheres and three polyhedra (cube, tetrahedron, and hexagonal prism) by performing NVT simulations over a series of volumes, iteratively scaling each box dimension by 0.95 and running 100,000 steps ($10\tau$) for equilibration followed by 100,000 steps ($10\tau$) for production at each volume. Each simulation consists of 125 particles with initially random positions and orientations. The diameter of the sphere is set to 2, while the volumes of the cube and hexagonal prism are set to 8, and that of the tetrahedron is 27. A skin layer thickness of $R^* = 0.15$ is applied, so the effective volume $v_o^*$ is taken as the Minkowski sum of the polyhedron and a sphere of radius $R^*$. The larger volume chosen for the tetrahedron reduces the error between the perfect polyhedron and the offset polyhedron with rounded corners, which is more pronounced for the tetrahedron than for the other two polyhedra. The system temperature is set to $T^* = 1.5$. Note that spherical particles are treated as having only three translational degrees of freedom, since self-rotation is not considered in this case, whereas other particles have six degrees of freedom. Therefore, the pressure \(P^*\) is given by~\cite{thompson2009general}
\begin{equation}
P^* = \frac{N_{\rm dof} T^*}{3V^*} + \frac{1}{3V^*} \sum_{i=1}^{N} \sum_{j>i}^{N} \mathbf{r}^*_{ij} \cdot \mathbf{f}^*_{ij},
\end{equation}
where \(N_{\rm dof}\) is the total number of degrees of freedom, \(\mathbf{r}^*_{ij} = \mathbf{r}^*_i - \mathbf{r}^*_j\) is the displacement vector between particles \(i\) and \(j\), and \(\mathbf{f}^*_{ij}\) is the force exerted on particle \(i\) by particle \(j\).

\begin{figure}[htbp]
    \centering
    \includegraphics[width=\linewidth]{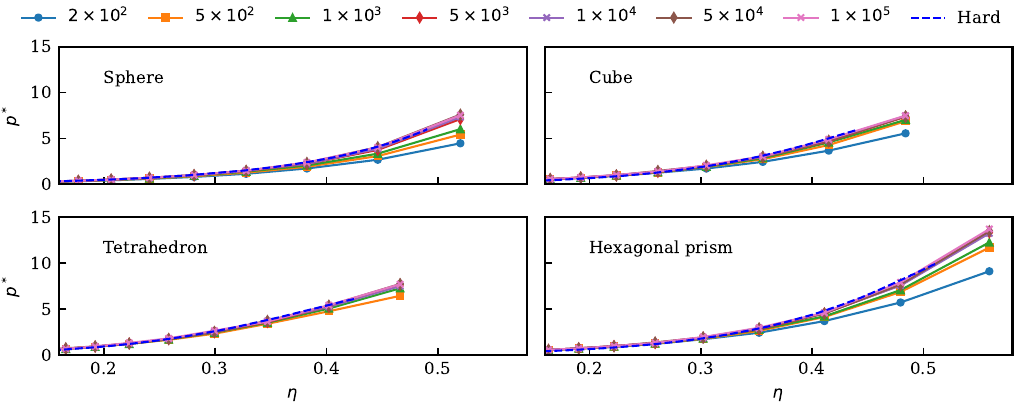}
    \caption{Reduced pressure \(p^*\) as a function of packing fraction \(\eta\) for spheres, cubes, tetrahedra, and hexagonal prisms at various values of the repulsive stiffness \(k_n^*\) (see Eq.~\eqref{eq:contact_f}). The case of \(k_n^* = 2 \times 10^2\) for tetrahedra was found to be unstable and is therefore omitted from the plot. The hard-particle reference data are taken from MC simulations reported by Irrgang \textit{et al.}~\cite{irrgang2017virial} As \(k_n^*\) increases, the reduced pressure curves progressively approach the hard-particle results.}
    \label{fig:eos}
\end{figure}

Figure~\ref{fig:eos} shows the reduced pressure, \(p^* = P^* v_o^* / T^*\), as a function of packing fraction for spheres, cubes, tetrahedra, and hexagonal prisms at various repulsive stiffness values \(k_n^*\) (see Eq.~\eqref{eq:contact_f}). The hard-particle reference data are taken from MC simulations reported by Irrgang \textit{et al.}~\cite{irrgang2017virial} The reduced pressure curves closely match the reference data at high $k_n$, confirming the accuracy of our contact definitions and demonstrating the reliability of our simulation framework for modeling complex convex particle systems. Moreover, calculating the equation of state for fluids of highly nonspherical bodies provides fundamental insight into the mechanistic role of particle shape in governing macroscopic thermodynamic properties, quantitatively linking shape to pressure, density, and packing, and revealing systematic trends that enable predictive understanding of nonspherical particle fluids.

\section{Conclusion and Future Outlook}
\label{sec:con}

Simulations of nonspherical rigid-body particles remain a powerful tool for exploring the dynamics of phenomena beyond the microscopic, such as colloidal systems. Maintaining accurate energy conservation in these simulations is essential, as it ensures that key dynamical properties can be reliably determined from a statistical mechanical perspective. In this paper, we have presented the framework and rigorously validated its applicability for simulating mixtures of arbitrary anisotropic particles. The implementation of vertex–boundary interactions for 2D particles, along with vertex–surface and edge–edge interactions for 3D particles, ensures smooth contact force generation and prevents particle intersection. Key properties, including packing, diffusivity, and the equation of state, are evaluated to demonstrate the capabilities of the framework, offering dynamic insights and nonequilibrium perspectives that extend beyond the reach of conventional MC simulations.

This framework can be further extended to a wide range of applications. One example is the study of scattering problems using simulation trajectories, along with the calculation of structural factors and the analysis of collective behaviors in nonspherical particle systems. Accurately extracting structural factors enables characterization of the spacing and orientation of building blocks in an assembly, as well as large-scale features that are otherwise intractable for interpretation, complementing experimental observations.\cite{heil2023computational, bryant2024advanced, chiang2025efficient} Another research direction involves investigating the complexity and self-assembly of biomolecules with intricate shapes and heterogeneous surface groups to design responsive materials and molecular machines.\cite{kahn2025encoding,teng2025macroscale,zhang2025design} To achieve this, future developments of the framework should enable users to specify points on a particle where forces or interactions can be applied to represent surface heterogeneity. Finally, this framework can be employed to systematically study hydrodynamic and lubrication properties in nonspherical particle systems, capturing how local contact geometry, tangential forces, and surface anisotropy couple with Brownian forces and torques to influence momentum transfer, rotational dynamics, and emergent collective flow under confined or dense conditions.\cite{kraft2013brownian,sanjeevi2018drag,heo2024colloidal,lee2024defect} Overall, the inherent dynamical fidelity of this framework provides a versatile tool for the colloid and granular communities to investigate the structural, dynamical, and hydrodynamic behaviors of complex nonspherical particle systems, enabling studies that probe deeply into nonequilibrium processes.

\section*{Acknowledgments}
H.S. and G.K.S. were supported by the U.S. Department of Energy (DOE), Office of Science, Basic Energy Sciences (BES), the Chemical Sciences, Geosciences, and Biosciences Division, Chemical Physics and Interfacial Sciences Program, FWP 16249. J.C. was supported by the U.S. DOE, Office of Science, BES, under the Synthesis and Processing Science Program, FWP 67554. C.J.M. was supported by the U.S. DOE, Office of Science, BES, Energy Frontier Research Center -- The Center for the Science of Synthesis Across Scales (CSSAS) located at the University of Washington (UW), award number DE-SC0019288. The authors sincerely thank Benjamin A. Helfrecht for his exploratory contributions and insightful discussions that greatly aided the development of this work.

\section*{Author Declarations}
\subsection*{Conflict of Interest}
The authors have no conflicts to disclose.

\subsection*{Author Contributions}
\noindent \textbf{Haoyuan Shi:} Conceptualization (equal); Formal analysis (lead); Investigation (lead); Methodology (lead); Software (lead); Validation (lead); Visualization (lead); Writing -- original draft (lead); Writing -- review \& editing (equal). \textbf{Christopher J. Mundy:} Conceptualization (equal); Formal analysis (equal); Funding acquisition (equal); Methodology (equal); Project administration (lead); Supervision (equal); Writing -- review \& editing (equal). \textbf{Gregory K. Schenter:} Conceptualization (equal); Formal analysis (equal); Funding acquisition (lead); Methodology (equal); Project administration (lead); Supervision (equal); Writing -- review \& editing (equal). \textbf{Jaehun Chun:} Conceptualization (equal); Formal analysis (equal); Funding acquisition (equal); Methodology (equal); Project administration (lead); Supervision (lead); Writing -- review \& editing (equal).

\section*{Data Availability Statement}

The source codes and simulation scripts supporting this study are openly available on GitHub at: \\
\url{https://github.com/Haoyuan-Shi/energy-conserving-contact-dynamics}

\bibliographystyle{unsrt}  
\bibliography{references}  

\begin{thebibliography}{10}

\bibitem{glotzer2007anisotropy}
Sharon~C Glotzer and Michael~J Solomon.
\newblock Anisotropy of building blocks and their assembly into complex structures.
\newblock {\em Nature materials}, 6(8):557--562, 2007.

\bibitem{damasceno2012predictive}
Pablo~F Damasceno, Michael Engel, and Sharon~C Glotzer.
\newblock Predictive self-assembly of polyhedra into complex structures.
\newblock {\em Science}, 337(6093):453--457, 2012.

\bibitem{li2012direction}
Dongsheng Li, Michael~H Nielsen, Jonathan~RI Lee, Cathrine Frandsen, Jillian~F Banfield, and James~J De~Yoreo.
\newblock Direction-specific interactions control crystal growth by oriented attachment.
\newblock {\em Science}, 336(6084):1014--1018, 2012.

\bibitem{de2015crystallization}
James~J De~Yoreo, Pupa~UPA Gilbert, Nico~AJM Sommerdijk, R~Lee Penn, Stephen Whitelam, Derk Joester, Hengzhong Zhang, Jeffrey~D Rimer, Alexandra Navrotsky, Jillian~F Banfield, et~al.
\newblock Crystallization by particle attachment in synthetic, biogenic, and geologic environments.
\newblock {\em Science}, 349(6247):aaa6760, 2015.

\bibitem{noid2023perspective}
William~George Noid.
\newblock Perspective: Advances, challenges, and insight for predictive coarse-grained models.
\newblock {\em The Journal of Physical Chemistry B}, 127(19):4174--4207, 2023.

\bibitem{li2023nanoparticle}
Dongsheng Li, Qian Chen, Jaehun Chun, Kristen Fichthorn, James De~Yoreo, and Haimei Zheng.
\newblock Nanoparticle assembly and oriented attachment: correlating controlling factors to the resulting structures.
\newblock {\em Chemical reviews}, 123(6):3127--3159, 2023.

\bibitem{heo2024colloidal}
Jaeyoung Heo, Pravalika Butreddy, Gregory~K Schenter, Christopher~J Mundy, James~J De~Yoreo, Elias Nakouzi, Jaewon Lee, and Jaehun Chun.
\newblock Colloidal phenomena reflect the interplay between interfacial solution structure, interparticle forces, and dynamical response.
\newblock {\em Current Opinion in Colloid \& Interface Science}, page 101887, 2024.

\bibitem{krim2024fundamental}
Jacqueline Krim and Alex~I Smirnov.
\newblock Fundamental mechanisms underlying the effectiveness of nanoparticle additives to lubricants: 25 examples linking nano-to macroscale friction.
\newblock {\em Lubricants}, 12(6):225, 2024.

\bibitem{shi2025incorporating}
Haoyuan Shi, Christopher~J Mundy, Gregory~K Schenter, and Jaehun Chun.
\newblock Incorporating the molecular-scale into a hydrodynamic description of confined aqueous systems.
\newblock {\em The Journal of Chemical Physics}, 163(13), 2025.

\bibitem{berardi1998gay}
Roberto Berardi, Carlo Fava, and Claudio Zannoni.
\newblock A gay--berne potential for dissimilar biaxial particles.
\newblock {\em Chemical physics letters}, 297(1-2):8--14, 1998.

\bibitem{berardi2008field}
Roberto Berardi, Luca Muccioli, and Claudio Zannoni.
\newblock Field response and switching times in biaxial nematics.
\newblock {\em The Journal of chemical physics}, 128(2), 2008.

\bibitem{nguyen2019aspherical}
Trung~Dac Nguyen and Steven~J Plimpton.
\newblock Aspherical particle models for molecular dynamics simulation.
\newblock {\em Computer Physics Communications}, 243:12--24, 2019.

\bibitem{ramasubramani2020mean}
Vyas Ramasubramani, Thi Vo, Joshua~A Anderson, and Sharon~C Glotzer.
\newblock A mean-field approach to simulating anisotropic particles.
\newblock {\em The Journal of Chemical Physics}, 153(8), 2020.

\bibitem{fraige2008vibration}
Feras~Y Fraige, Paul~A Langston, Andrew~J Matchett, and John Dodds.
\newblock Vibration induced flow in hoppers: Dem 2d polygon model.
\newblock {\em Particuology}, 6(6):455--466, 2008.

\bibitem{wang2011particle}
J~Wang, HS~Yu, P~Langston, and F~Fraige.
\newblock Particle shape effects in discrete element modelling of cohesive angular particles.
\newblock {\em Granular Matter}, 13(1):1--12, 2011.

\bibitem{thompson2022lammps}
Aidan~P Thompson, H~Metin Aktulga, Richard Berger, Dan~S Bolintineanu, W~Michael Brown, Paul~S Crozier, Pieter~J In't~Veld, Axel Kohlmeyer, Stan~G Moore, Trung~Dac Nguyen, et~al.
\newblock Lammps-a flexible simulation tool for particle-based materials modeling at the atomic, meso, and continuum scales.
\newblock {\em Computer physics communications}, 271:108171, 2022.

\bibitem{spellings2017gpu}
Matthew Spellings, Ryan~L Marson, Joshua~A Anderson, and Sharon~C Glotzer.
\newblock Gpu accelerated discrete element method (dem) molecular dynamics for conservative, faceted particle simulations.
\newblock {\em Journal of Computational Physics}, 334:460--467, 2017.

\bibitem{anderson2020hoomd}
Joshua~A Anderson, Jens Glaser, and Sharon~C Glotzer.
\newblock Hoomd-blue: A python package for high-performance molecular dynamics and hard particle monte carlo simulations.
\newblock {\em Computational Materials Science}, 173:109363, 2020.

\bibitem{de2022spiers}
James~J De~Yoreo, Elias Nakouzi, Biao Jin, Jaehun Chun, and Christopher~J Mundy.
\newblock Spiers memorial lecture: Assembly-based pathways of crystallization.
\newblock {\em Faraday Discussions}, 235:9--35, 2022.

\bibitem{van2014understanding}
Greg van Anders, Daphne Klotsa, N~Khalid Ahmed, Michael Engel, and Sharon~C Glotzer.
\newblock Understanding shape entropy through local dense packing.
\newblock {\em Proceedings of the National Academy of Sciences}, 111(45):E4812--E4821, 2014.

\bibitem{van2015digital}
Greg Van~Anders, Daphne Klotsa, Andrew~S Karas, Paul~M Dodd, and Sharon~C Glotzer.
\newblock Digital alchemy for materials design: Colloids and beyond.
\newblock {\em Acs Nano}, 9(10):9542--9553, 2015.

\bibitem{cersonsky2018relevance}
Rose~K Cersonsky, Greg van Anders, Paul~M Dodd, and Sharon~C Glotzer.
\newblock Relevance of packing to colloidal self-assembly.
\newblock {\em Proceedings of the National Academy of Sciences}, 115(7):1439--1444, 2018.

\bibitem{tahmasebi2023state}
Pejman Tahmasebi.
\newblock A state-of-the-art review of experimental and computational studies of granular materials: Properties, advances, challenges, and future directions.
\newblock {\em Progress in materials science}, 138:101157, 2023.

\bibitem{agarwal2011mesophase}
Umang Agarwal and Fernando~A Escobedo.
\newblock Mesophase behaviour of polyhedral particles.
\newblock {\em Nature materials}, 10(3):230--235, 2011.

\bibitem{ramasubramani2021coxeter}
Vyas Ramasubramani, Bradley Dice, Tobias Dwyer, and Sharon Glotzer.
\newblock coxeter: A python package for working with shapes.
\newblock {\em Journal of Open Source Software}, 6(63), 2021.

\bibitem{stukowski2009visualization}
Alexander Stukowski.
\newblock Visualization and analysis of atomistic simulation data with ovito--the open visualization tool.
\newblock {\em Modelling and simulation in materials science and engineering}, 18(1):015012, 2009.

\bibitem{martyna1994constant}
Glenn~J Martyna, Douglas~J Tobias, and Michael~L Klein.
\newblock Constant pressure molecular dynamics algorithms.
\newblock {\em J. chem. Phys}, 101(4177):10--1063, 1994.

\bibitem{torquato2010jammed}
Salvatore Torquato and Frank~H Stillinger.
\newblock Jammed hard-particle packings: From kepler to bernal and beyond.
\newblock {\em Reviews of modern physics}, 82(3):2633--2672, 2010.

\bibitem{larsen2016robust}
Peter~Mahler Larsen, S{\o}ren Schmidt, and Jakob Schi{\o}tz.
\newblock Robust structural identification via polyhedral template matching.
\newblock {\em Modelling and Simulation in Materials Science and Engineering}, 24(5):055007, 2016.

\bibitem{thompson2009general}
Aidan~P Thompson, Steven~J Plimpton, and William Mattson.
\newblock General formulation of pressure and stress tensor for arbitrary many-body interaction potentials under periodic boundary conditions.
\newblock {\em The Journal of chemical physics}, 131(15), 2009.

\bibitem{irrgang2017virial}
M~Eric Irrgang, Michael Engel, Andrew~J Schultz, David~A Kofke, and Sharon~C Glotzer.
\newblock Virial coefficients and equations of state for hard polyhedron fluids.
\newblock {\em Langmuir}, 33(42):11788--11796, 2017.

\bibitem{heil2023computational}
Christian~M Heil, Yingzhen Ma, Bhuvnesh Bharti, and Arthi Jayaraman.
\newblock Computational reverse-engineering analysis for scattering experiments for form factor and structure factor determination (“p (q) and s (q) crease”).
\newblock {\em JACS Au}, 3(3):889--904, 2023.

\bibitem{bryant2024advanced}
Gary Bryant, Amani Alzahrani, Saffron~J Bryant, Reece Nixon-Luke, Jitendra Mata, and Rohan Shah.
\newblock Advanced scattering techniques for characterisation of complex nanoparticles in solution.
\newblock {\em Advances in Colloid and Interface Science}, 334:103319, 2024.

\bibitem{chiang2025efficient}
Huat~Thart Chiang, Zhiyin Zhang, Kiran Vaddi, F~Akif Tezcan, and Lilo~D Pozzo.
\newblock Efficient analysis of small-angle scattering curves for large biomolecular assemblies using monte carlo methods.
\newblock {\em Applied Crystallography}, 58(3), 2025.

\bibitem{kahn2025encoding}
Jason~S Kahn, Brian Minevich, Aaron Michelson, Hamed Emamy, Jiahao Wu, Huajian Ji, Alexia Yun, Kim Kisslinger, Shuting Xiang, Nanfang Yu, et~al.
\newblock Encoding hierarchical 3d architecture through inverse design of programmable bonds.
\newblock {\em Nature materials}, pages 1--10, 2025.

\bibitem{teng2025macroscale}
Feiyue Teng, Honghu Zhang, Dmytro Nykypanchuk, Ruipeng Li, Lin Yang, Nikhil Tiwale, Zhaoyi Xi, Mingzhao Liu, Mingxin He, Shuai Zhang, et~al.
\newblock Macroscale-area patterning of three-dimensional dna-programmable frameworks.
\newblock {\em Nature communications}, 16(1):3238, 2025.

\bibitem{zhang2025design}
Zhiyin Zhang, Huat~T Chiang, Ying Xia, Nicole Avakyan, Ravi~R Sonani, Fengbin Wang, Edward~H Egelman, James~J De~Yoreo, Lilo~D Pozzo, and F~Akif Tezcan.
\newblock Design of light-and chemically responsive protein assemblies through host-guest interactions.
\newblock {\em Chem}, 2025.

\bibitem{kraft2013brownian}
Daniela~J Kraft, Raphael Wittkowski, Borge Ten~Hagen, Kazem~V Edmond, David~J Pine, and Hartmut L{\"o}wen.
\newblock Brownian motion and the hydrodynamic friction tensor for colloidal particles of complex shape.
\newblock {\em Physical Review E—Statistical, Nonlinear, and Soft Matter Physics}, 88(5):050301, 2013.

\bibitem{sanjeevi2018drag}
Sathish~KP Sanjeevi, JAM Kuipers, and Johan~T Padding.
\newblock Drag, lift and torque correlations for non-spherical particles from stokes limit to high reynolds numbers.
\newblock {\em International Journal of Multiphase Flow}, 106:325--337, 2018.

\bibitem{lee2024defect}
Jaewon Lee, Zexi Lu, Zhigang Wu, Colin Ophus, Gregory~K Schenter, James~J De~Yoreo, Jaehun Chun, and Dongsheng Li.
\newblock Defect self-elimination in nanocube superlattices through the interplay of brownian, van der waals, and ligand-based forces and torques.
\newblock {\em ACS nano}, 18(47):32386--32400, 2024.

\end{thebibliography}

\end{document}